\documentclass[twocolumn,showpacs,aps,amssymb,floatfix,prd,amsmath,preprintnumbers]{revtex4} 
\usepackage{epstopdf}
\usepackage{capt-of}
\usepackage{graphicx}  
\usepackage{dcolumn}   
\usepackage{bm}
\begin{document}

\def \reals{{\mathbb R}}
\def\be{\begin{equation}}
\def\ee{\end{equation}}
\def\bea{\begin{eqnarray}}
\def\eea{\end{eqnarray}}
\def\nn{\nonumber}
\def\th{\theta}
\def\ph{\phi}
\def\lt{\left}
\def\rt{\right}
\def\ed{\end{document}}
\def\degree{\mathop{\rm {{}^\circ}}}
\input epsf.tex
\title{Negative Time Delay in Strongly Naked Singularity Lensing}

\author{Justin P.~DeAndrea and Kevin M.~Alexander}
     \affiliation{Department of Physics, The College of New Jersey, 2000 Pennington   
                   Rd., Ewing, NJ 08628, USA.}

\begin{abstract}
We model the supermassive galactic center of the Milky Way galaxy as a strongly naked singularity lens described by the Janis-Newman-Winicour metric. This metric has an ordinary mass and massless scalar charge parameters. For very accurate results, we use the Virbhadra-Ellis lens equation for computations. The galactic center serving as gravitational lens gives rise to 4 images: 2 images on the same side as the source and 2 images on the opposite side of the source from the optic axis. We compute positions and time delays of these images for many values of the angular source position. The time delays of primary images decrease with an increase in angular source position and are always negative. The time delays of the other 3 images are negative for small angular source position; however, they increase with an increase in angular source position. Such observations would support strongly naked singularity interpretation of the galactic center and, if ever observed, would disprove the cosmic censorship hypothesis proposed by Roger Penrose as well as a weaker version by Virbhadra that allows existence of weakly, but not marginally and strongly naked singularities.

\end{abstract}



\maketitle


\section{Introduction}

Either convincing proof or disproof of the Cosmic Censorship Hypothesis (CCH) developed by Penrose\cite{pen69} is a compelling unsolved problem in classical general relativity. This hypothesis generally states that spacetime singularities of physically realistic gravitational collapse are hidden from an observer by the event horizon of a black hole\cite{pen69,pen70}. Thus, by definition, this hypothesis implies that naked (visible) singularities cannot exist in a realistic gravitational collapse\cite{pen98}. A complete proof or disproof for the CCH has not been found, despite extensive research (for details, see in \cite{wald97}. Providing a counter-example to the CCH could be of great importance towards understanding the physical laws of quantum gravity. However, our current observational, analytic, and numerical techniques are not yet sufficiently refined to produce a proof or disproof of the cosmic censorship hypothesis\cite{pen98}. In light of this, there is now a strong need to explore whether or not this hypothesis could be tested observationally. Virbhadra and his collaborators have done pioneering work in this field. Their study of gravitational lensing shows great potential for the differentiation of black holes from naked singularities based off qualitatively different lensing features\cite{vnc98,ve00,cve01,ve02,vk08,v09}. Inspired by their outbreaking results on one of the most important unsolved problems in relativistic astrophysics, a large number of researchers have studied gravitational lensing by different kinds of black holes and naked singularities (see \cite{per1,per2,per3,zak,many1,many2} and references therein.) \\
\indent
Null geodesics in the Schwarzschild space-time were solved for the first time in 1931\cite{hag31}. Lensing due to a strong gravitational field of a black hole was not studied until it was done by Darwin in 1958 and 1961\cite{dar58}. Darwin called these images {\em ghosts}. There was no gravitational lens equation for strong gravitational field lensing and therefore Darwin's analysis was not very accurate. In 2000, Virbhadra and Ellis \cite{ve00} gave a new lens equation that allows arbitrarily large as well as small deflection angles of light and this outbreaking work resurrected strong field gravitational lensing studies. Being unaware of Darwin's work, Virbhadra and Ellis coined strong field lensing images
(formed due to deflection angle $\hat{\alpha}> 3\pi/2$) {\em relativistic images}. Now, instead of being referred to as ghost images, most researchers call these relativistic images. Thereafter, a few other lens equations have been proposed. However, the one given by Virbhadra and Ellis is the simplest in form, easiest to use, and gives very accurate results; this equation is now well-known as {\em Virbhadra-Ellis lens equation}. Perlick commented that this gravitational lens equation is almost exact, \cite{per2} and it has found the biggest resonance in the literature\cite{per3}.
\indent
Based on existence of event horizon and photon spheres, Virbhadra and his collaborators\cite{ve02,vk08} identified the possibility of four different types of singularities: black hole singularities (BHS), weakly naked singularities (WNS), marginally strongly naked singularities (MSNS), and strongly naked singularities (SNS). It was concluded that BHS, WNS, and MSNS have qualitatively similar lensing behavior and therefore are very difficult to differentiate by their slightly differing qualitative features until we have developed very advanced level observational techniques. However, they
found that lensing by SNS is qualitatively quite different. This distinction allows for a critical test for the existence of a SNS.
In\cite{vk08}, it was also concluded that SNS lensing would be a more efficient cosmic telescope than black hole lensing, because it gives rise to images of smaller time delay and higher magnification. Our research focused on lensing due to a SNS. Virbhadra and Keeton\cite{vk08} studied time delay and magnification centroid of images due to black holes, WNS, MSNS, and SNS. They found that in gravitational lensing due to a SNS, negative time delays could be plausible. Observing such time delays for studied values of angular source position will prove the existence of naked singularities, and thus disprove the CCH.
\indent
In this work, we study time delays of images due to gravitational lensing by a particular SNS in a great detail. Given the lack of scientific evidence in support of the CCH, there is no reason to not model the supermassive dark objects at centers of galaxies as SNS. Therefore, it is reasonable that we study the time delay due to gravitational lensing by the Milky Way galactic center modeled as a SNS. Light propagation is determined through a static spherically symmetric Janis-Newman-Winicour (JNW) metric \cite{jnw68}. All computations have been performed using the MATHEMATICA software\cite{wolf}.\\


\section{\label{sec:LE&DE}Virbhadra-Ellis Lens Equation and the Janis-Newman-Winicour naked Singularity Spacetime}

In this  section, we briefly discuss the equations and results obtained by Virbhadra {\it et al.}  required to determine  positions and time delays of images associated with  gravitational  lensing. We use  the same symbols as used by them.
\indent
The Virbhadra-Ellis gravitational lens equation\cite{ve00}, which permits small as well as large bending angles of light, is 
\begin{equation}
\tan\beta = \tan\theta - \alpha
\end{equation}
with
\begin{equation}
\alpha \equiv \frac{D_{ds}}{D_s} [\tan\theta + \tan(\hat{\alpha} - \theta)].
\end{equation}

$D_{s}$ is the observer-source distance, $D_{ds}$ is the lens-source distance, and $D_d$  is the observer-lens distance. $\hat{\alpha}$  is the light bending angle. $\theta$  and $\beta$  are, respectively, angular positions of an image and unlensed source measured from the optical axis. The impact parameter $J = D_d \sin \theta$. 

\indent
Virbhadra {\it et al}.\cite{vnc98} considered the metric
\begin{equation}
ds^2 = B(r)dt^2 - A(r)dr^2 - D(r)r^2(d\vartheta + \sin^2\vartheta d\phi^2)
\end{equation}
This is a general static and spherically symmetric metric. They calculated the deflection $\hat{\alpha}\left(r_o\right)$ and impact parameter $J\left(r_0\right)$  for a light ray with the closest distance of approach $r_0$. These are expressed by

\begin{widetext}
\begin{equation}
    \hat{\alpha}\lt(r_0\rt)
              = 2 {\int_{r_0}}^{\infty}\lt(\frac{A(r)}{D(r)}\rt)^{1/2}
                       \lt[
                       \lt(\frac{r}{r_0}\rt)^2\frac{D(r)}{D(r_0)}\frac{B(r_0)}{B(r)}-1
                    \rt]^{-1/2} \frac{dr}{r}   - \pi
      \label{GenAlphaHat}
\end{equation}

and

\begin{equation}
    J\lt(r_0\rt)  = r_0  \sqrt{\frac{D(r_0)}{B(r_0)}} \text{.}
\end{equation}
\end{widetext}

\indent
The most general static and spherically symmetric solution to the Einstein massless scalar field equations initially obtained by Janis, Newman and Winicour \cite{jnw68} (characterized by constant and real parameters, the mass $M$  and scalar charge $q$) was later  re-expressed by Virbhadra \cite{v97} in a more convenient  form:
\begin{eqnarray}
   ds^2 &=&   \left(1-\frac{b}{r}\right)^{\nu} dt^2
      - \left(1-\frac{b}{r}\right)^{-\nu} dr^2 \nonumber \\
      &-& \left(1-\frac{b}{r}\right)^{1-\nu}  
      r^2 \left(d\vartheta^2  +\sin^2\vartheta \  d\varphi^2\right) 
    \label{Janis-Newman-WinicourMetric} 
\end{eqnarray}
and the massless scalar field
\begin{equation}
\Phi = \frac{q}{b\sqrt{4\pi}}\ln\left(1-\frac{b}{r}\right),
\end{equation}
with
\begin{equation}
\nu = \frac{2M}{b}\; \mbox {and}\;  b = 2\sqrt{M^2 + q^2}.
\end{equation}
The Janis-Newman-Winicour spacetime has only one photon sphere situated at the radial distance\cite{cve01,ve02,vk08}
\begin{equation}
r_{ps}=\frac{b(1+2\nu)}{2}.
\end{equation}
The photon sphere exists only for $1/2<\nu\leq1$. Virbhadra {\it et al}. \cite{vjj97} showed that the JNW metric has a strong curvature globally naked singularity at $r=b$.

 Defining
\begin{equation}
\rho=\frac{r}{b} \;\;  \mbox{and} \; \;  \rho_0=\frac{r_0}{b},
\end{equation}
the deflection angle $\hat{\alpha}$ for a light ray in the Janis-Newman-Winicour spacetime is expressed in the form\cite{vnc98,ve02,vk08}
\begin{widetext}
\begin{equation}
\hat{\alpha}(\rho_0)=2\int_{\rho_0}^{\infty}\frac{d\rho}{\rho\sqrt{1-\frac{1}{\rho}}\sqrt{(\frac{\rho}{\rho_0})^2\left(1-\frac{1}{\rho}\right)^{1-2\nu}\left(1-\frac{1}{\rho_0}\right)^{2\nu-1}-1}}-\pi.
\end{equation}
\\*
\\*
\\*
\\*
\\*
\\*
\\*
\\*

\end{widetext}

\begin{center}
\includegraphics[width=3in,height=3in]{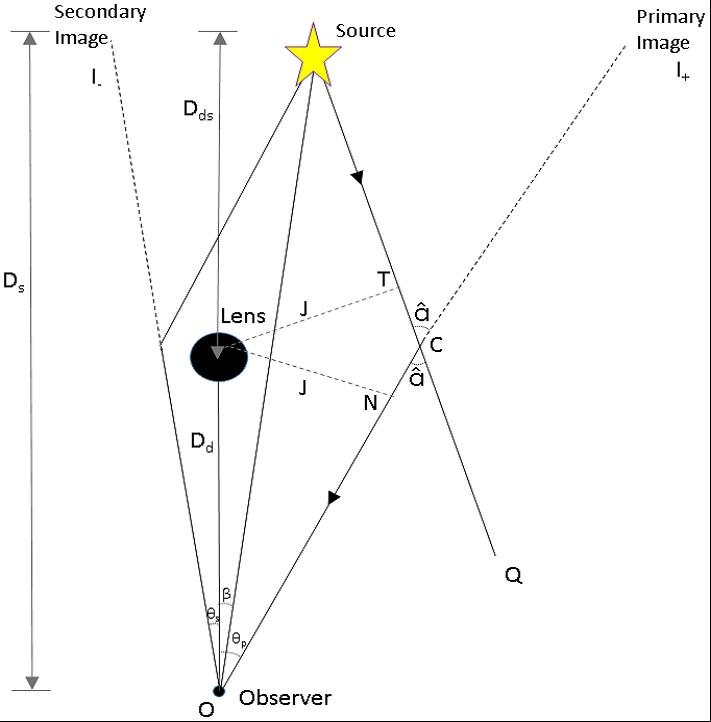}
\DeclareGraphicsExtensions{.png}
\captionof{figure}{Schematic diagram of  gravitational lensing. All angles and distances presented in the lens equation are shown. }
\label{Fig1} %
\end{center}


Virbhadra and Keeton \cite{vk08} calculated time delay that a light ray suffers in a general static spherically symmetric  spacetime. Then, they used that to  find  the time delay in the JNW spacetime and that 
given below:

\begin{equation}
  \tau\lt(\rho_0\rt) 
      = \frac{2M}{\nu} 
         \lt[         
             {\int_{\rho_0}}^{{\cal X}_s} \frac{d\rho}{f\lt(\rho\rt)} + {\int_{\rho_0}}^{{\cal X}_o} \frac{d\rho}{f\lt(\rho\rt)}          
          \rt]  
       - D_s \sec\beta                                        
\end{equation}
with
\begin{eqnarray}
    {\cal X}_s &=& \frac{\nu}{2} \frac{D_s}{M} \sqrt{   \lt(\frac{D_{ds}}{D_s}\rt)^2+\tan^2\beta  } \text{,}\nonumber\\
    {\cal X}_o &=& \frac{\nu}{2} \frac{D_d}{M} \text{,} 
\end{eqnarray}
and
\begin{equation}
    f\lt(\rho\rt)={ \sqrt{ \lt(1-\frac{1}{\rho}\rt)^{2\nu}-\lt(\frac{\rho_0}{\rho}\rt)^2\lt(1-\frac{1}{\rho}\rt)^{4\nu-1}\lt(1-\frac{1}{\rho_0}\rt)^{1-2\nu}}}  \text{.}                             
    \label{f}
\end{equation}
The differential time delay $\Delta\tau$ of an image with time delay $\tau$ is measured in reference to the direct image, and  is expressed by

\begin{equation}
\Delta\tau=\tau-\tau_p,
\end{equation}
where $\tau_p$ is the time delay of the primary image.

\section{Computations and results}

Virbhadra and Keeton\cite{vk08} studied qualitatively 3 distinct types of SNS lensing. In the first ($\nu = 0.04$), there are 4 images. The primary image has positive time delays for small values of $\beta$ and negative time delays for large values of $\beta$. The remaining 3 images have positive time delays for all values of $\beta$. For the second type ($\nu = 0.02$), there are still 4 images. However, the primary image has negative time delay for all values of $\beta$. Time delays of the other 3 images have negative, zero, and positive time delays depending on the angular source position. In the third type of SNS lensing ($\nu = 0.001$), there are only 2 images on the same side as the source. Both have negative time delays. The time delay of the primary image decreases as $\beta$ increases, whereas the time delay of the inner image increases. When there is a perfect alignment of source, types 1 and 2 SNS lensing gives double Einstein rings; however type 3 gives no images at all.\\
\indent
In the present article, in order to study the lensing behavior of SNS between types 2 and 3, we examine a $\nu = 0.01$ case. We consider the Galactic supermassive center as a SNS ($\nu$ = 0.01, mass $M$ = 3.61 x $10^6 M_{\bigodot}$, $D_d$ = 7.62 kpc, $\frac{D_{ds}}{D_s} = \frac{1}{2}$) and, using MATHEMATICA, compute image positions, time delays, and differential time delays of images for a large number of angular source positions. We model the galactic MDO as JNW SNS lens. JNW spacetime has been recently proven to be stable under scalar field perturbations\cite{sa13}. Virbhdara and Ellis {\it et al}. \cite{ve02} determined that, in gravitational lensing by SNS, no relativistic images are produced. Therefore, the time delays for relativistic images are not considered in this article.\\
\indent
Table I displays the image positions, time delays, and differential time delays at various levels of $\beta$ for images on the opposite side of the source, while Table II displays the same lensing quantities for images on the same side of the source. Figures 2 and 3 show time delays and differential time delays of images, respectively, plotted against the angular position $\beta$ of the source. As differential time delay is considered in reference to the primary image, Fig. 3 has only 3 curves whereas Fig. 2 has 4 curves. For $\beta = 0$, there are 2 concentric Einstein rings. As $\beta$ increases, the angular distance between the images on the same side of the source increases: the primary image goes away from the optic axis whereas the inner image moves toward the optic axis. For the 2 images on the opposite side of the source, angular separation decreases: the outer and inner images move, respectively, toward and away from the optic axis. In addition, we get an exciting result that all 4 images have negative time delays even for $\beta = 0$. This was not found for types of SNS lensing studied by Virbhadra and Keeton \cite{vk08}. The time delay of the primary image decreases with increasing $\beta$. For other 3 images, time delay increases with
increasing $\beta$; however remains negative for a large value of $\beta$, as obvious on Fig. 2. Differential time delays of 3 images (1 on the same side of the source and 2 on opposite side of the source) are always positive and increase with increase in the value of $\beta$.

\begingroup
\squeezetable
\begin{table*}
\caption{\label{tab:Table1} \scriptsize  Image positions ($\theta$), time delays ($\tau$), and differential time delays ($\Delta\tau$) of  secondary image ($s$) and the inner image on the secondary side ($s1$) for various values of angular source position are tabulated. Angles are shown in  arcseconds and  times in minutes. The Galactic center is modeled as a Janis-Newman-Winicour strongly naked singularity $\nu = 0.01$. $M/D_d \approx 2.26 \times  10^{-11}$ and $D_{ds}/D_s=1/2$.  }
\begin{ruledtabular}
 \begin{tabular}{l|lllllll}
\multicolumn{1}{c|}{ $\beta$}&
\multicolumn{6}{c}{Images on the opposite side of the source.}\\
  &         $\theta_s$                &    $\tau_s $                   &   $\Delta\tau_s$                &   $\theta_{s1}$                & $\tau_{s1} $              & $\Delta\tau_{s1} $  \\
\hline
 \hline\noalign{\smallskip}
$0          $&$-1.383568            $&$-43.69854          $&$  0            $&$ -0.0091760  $&$  -39.523                     $&$ 4.175815$ \\    
$10^{-5}    $&$-1.383563            $&$ -43.69853          $&$ 0.000017     $&$ -0.0091760  $&$ -39.523                      $&$ 4.175824$   \\      
$10^{-3}    $&$-1.383066            $&$-43.69769          $&$0.001700       $&$ -0.0091760  $&$ -39.523                   $&$ 4.176671 $  \\      
$0.01    $&$ -1.378560              $&$-43.69002          $&$ 0.017003      $&$ -0.0091764  $&$  -39.523                    $&$ 4.184388$   \\       
$0.05          $&$-1.358709         $&$ -43.65564         $&$ 0.085021      $&$  -0.0091782 $&$  -39.522                    $&$ 4.218993$   \\       
$0.1         $&$ -1.334301          $&$ -43.61197        $&$  0.170070      $&$ -0.0091803  $&$   -39.519                    $&$  4.262955$   \\     
$1.0         $&$ -0.9692883         $&$-42.67715          $&$ 1.73653       $&$ -0.0092201  $&$-39.210                       $&$ 5.203846$ \\         
$1.5        $&$ -0.8209932          $&$-42.01896          $&$ 2.67003       $&$  -0.0092424 $&$ -38.823                     $&$5.866039 $ \\          
$2.0        $&$-0.7035052           $&$-41.24781          $&$ 3.67619       $&$  -0.0092650 $&$ -38.282                      $&$ 6.641556 $ \\           
$3.0        $&$-0.5356993           $&$ -39.33473          $&$  5.96999     $&$ -0.0093109  $&$ -36.741                      $&$8.564160 $ \\           
$4.0        $&$-0.4260236           $&$ -36.89084          $&$ 8.71160      $&$ -0.0093577  $&$  -34.584                    $&$11.01828 $ \\           
$6.0       $&$-0.2965909            $&$ -30.31079         $&$  15.7371      $&$  -0.0094542 $&$  -28.428                     $&$ 17.61997 $ \\          
$8.0          $&$-0.2247471         $&$ -21.39161          $&$ 24.9837      $&$ -0.0095548  $&$  -19.814                     $&$ 26.56169 $ \\           
$10.0   $&$-0.1796498               $&$-10.08453          $&$36.5486        $&$  -0.0096600 $&$   -8.7413                   $&$ 37.89185  $ \\         
 
\end{tabular}
\end{ruledtabular}
\end{table*}
\endgroup


\begingroup
\squeezetable
\begin{table*}
\caption{\label{tab:Table1} \scriptsize  Image positions ($\theta$) and  time delays ($\tau$) of   primary image ($p$) and the inner image on  the primary side ($p1$), and differential time delays ($\Delta\tau$) of the  inner image  on the primary side for various values of angular source position are tabulated. Angles are shown in  arcseconds and  times in minutes. The Galactic center is modeled as a Janis-Newman-Winicour strongly naked singularity {$\nu = 0.01$}. $M/D_d \approx 2.26 \times  10^{-11}$ and $D_{ds}/D_s=1/2$.  }

\begin{ruledtabular}
 \begin{tabular}{l|llllll}
\multicolumn{1}{c|}{ $\beta$}&
\multicolumn{6}{c}{Images on the same side of the source.}\\
  &         $\theta_{p1}$                &    $\tau_{p1} $                   &   $\Delta\tau_{p1}$                &   $\theta_{p}$                & $\tau_{p} $   \\
\hline
 \hline\noalign{\smallskip}
$0          $&$9.18\times10^{-3}           $&$-39.52272          $&$4.175815           $&$1.383568             $&$-43.69854           $&$   $ \\
$10^{-5}    $&$9.18\times10^{-3}           $&$-39.52272          $&$4.175823           $&$1.383573             $&$-43.69854           $&$   $ \\
$10^{-3}    $&$9.18\times10^{-3}             $&$-39.52273          $&$4.176660           $&$1.383070             $&$-43.69939           $&$   $ \\
$0.01    $&$9.18\times10^{-3}          $&$-39.52275          $&$4.18429           $&$1.383593             $&$-43.70702           $&$   $ \\
$0.05          $&$9.17\times10^{-3}             $&$-39.52223          $&$4.261827           $&$1.408876             $&$-43.74066           $&$   $ \\
$0.1         $&$9.17\times10^{-3}             $&$-39.52021          $&$4.261827           $&$1.434635             $&$-43.78204          $&$   $ \\
$1.0         $&$9.13\times10^{-3}             $&$-39.22111          $&$5.192569           $&$1.972436             $&$-44.41368           $&$   $ \\
$1.5        $&$9.11\times10^{-3}           $&$-38.83987          $&$5.849123           $&$2.325413             $&$-44.68899           $&$   $ \\
$2.0        $&$9.09 \times10^{-3}            $&$-38.30500          $&$6.619001           $&$2.708947             $&$-44.92400           $&$   $ \\
$3.0        $&$9.05 \times10^{-3}           $&$-38.77439          $&$8.530325           $&$3.542555             $&$-45.30472           $&$   $ \\
$4.0        $&$9.01\times10^{-3}           $&$-34.62929          $&$10.97316           $&$4.433728             $&$-45.60245           $&$   $ \\
$6.0       $&$8.93 \times10^{-3}         $&$-28.49558          $&$17.55228           $&$6.305180             $&$-46.04768           $&$   $ \\
$8.0          $&$8.85 \times10^{-3}           $&$-19.90390          $&$26.47139           $&$8.233777             $&$-46.37528           $&$   $ \\
$10.0   $&$8.78 \times10^{-3}          $&$-8.854225          $&$37.77891           $&$10.18896             $&$-46.63314           $&$   $ \\
\end{tabular}
\end{ruledtabular}
\end{table*}
\endgroup
\begin{center}
\includegraphics[width=3in,height=3in]{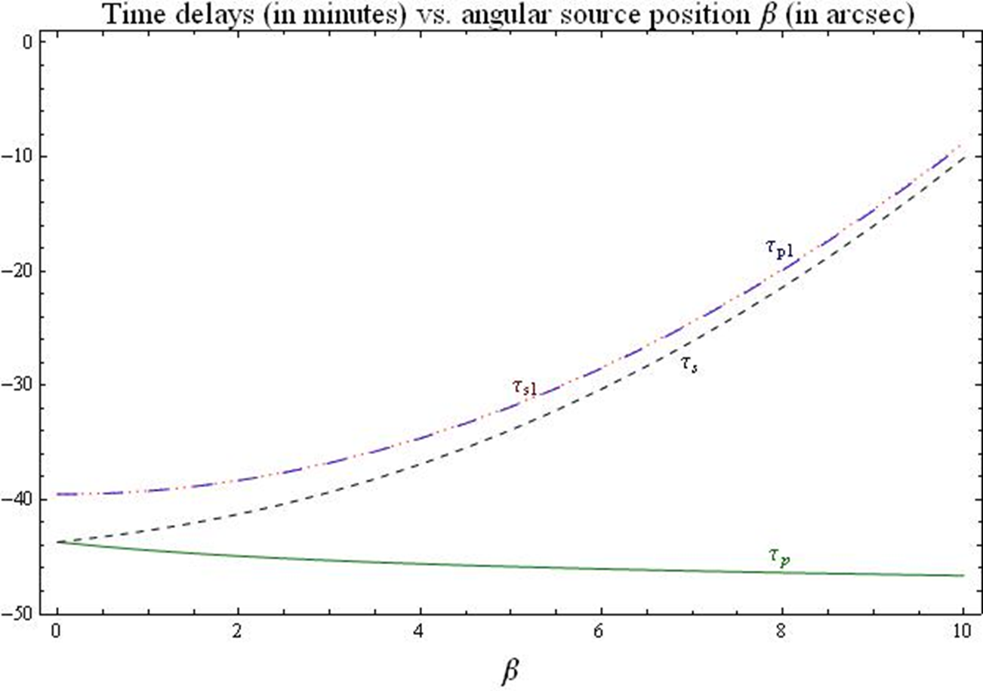}
\DeclareGraphicsExtensions{.png}
\captionof{figure}{{\tiny (color online)} \scriptsize This figure demonstrates variations of time delays of primary image $\tau_p$, secondary image $\tau_s$, inner image on primary side $\tau_{p1}$, and inner image on the secondary side $\tau_{s1}$ with change in angular source position $\beta$. Time delays and angular source positions are given, respectively, in minutes and arcseconds. The Galactic center is modeled as a strongly naked singularity with $\nu = 0.01$ and has  $M/D_d \approx 2.26 \times  10^{-11}$. We consider  $D_{ds}/D_s=1/2$. }
\label{Fig2} %
\end{center}

\begin{center}
\includegraphics[width=3in,height=3in]{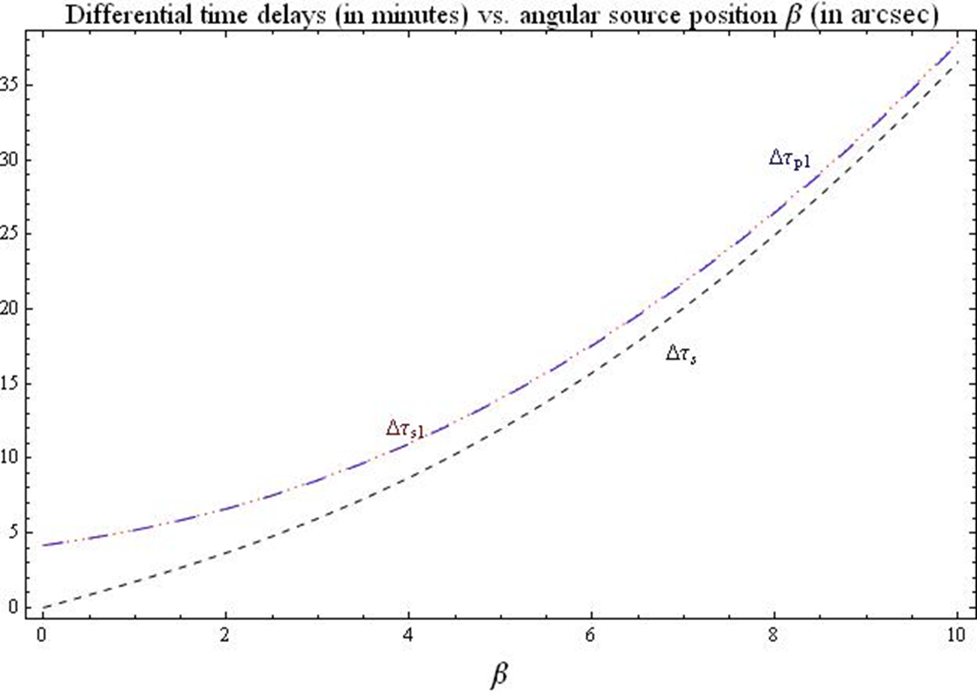}
\DeclareGraphicsExtensions{.png}
\captionof{figure}{{\tiny (color online)} \scriptsize  This figure demonstrates variations of  differential time delays of  secondary image $\Delta\tau_s$, inner image on primary side $\Delta\tau_{p1}$, and inner image on the secondary side $\Delta\tau_{s1}$ with change in angular source position $\beta$.  Differential time dealys and angular source positions are given, respectively, in minutes and arcseconds. The Galactic center is modeled as a strongly naked singularity with $\nu = 0.01$ and has  $M/D_d \approx 2.26 \times  10^{-11} $.  We consider  $D_{ds}/D_s=1/2$. }
\label{Fig3} %
\end{center}

\section{Summary and Discussion}

Virbhadra and Keeton\cite{vk08} studied gravitational lensing by three different kinds of Janis-Newman-Winicour SNS ($\nu$ = 0.04, $\nu$ = 0.02, and $\nu$ = 0.001). For the $\nu$ = 0.001 case, no Einstein ring is present when $\beta$ = 0, and no images appear on the opposite side of the source for large $\beta$. Both the $\nu$ = 0.04 and $\nu$ = 0.02 cases have double Einstein rings when $\beta$ = 0, which break into 4 images as $\beta$ increases. Virbhadra and Keeton found that as the value of $\nu$ decreases (that is, as the scalar charge is increased with the ADM mass $M$ fixed) the time delays became more negative \cite{vk08}. Virbhadra's initiative to study gravitational lensing by naked singularities arises from energy-momentum distributions in spacetimes (see \cite{energy} and references therein.) Before Virbhadra carried out computations, he was sure that JNW spacetime with large values of $\left(q/M\right)^2$ would work as divergent lenses and would therefore yield images with negative time delays, and his intuition came true\cite{vk08}. Virbhadra's seminal work on naked singularity lensing has been a source of inspiration for astrophysicists to work on this topic.\\
\indent
In this paper, we investigated a different type of Janis-Newman-Winicour SNS. We studied $\nu=0.01$ case and found a new exciting result. We found that, different from the types of SNS studied by Virbhadra and Keeton,
not only for the primary image, but also for the other 3 images, time delays are negative even for very small values of $\beta$. Most interestingly, time delays of double Einstein rings are also negative. The time delay of the primary image decreases and for the other 3 images increases as the source moves away from the optic axis. The primary image has negative time delays for any value of $\beta$. Time delays of other 3 images remain negative until a very large value of $\beta$. Figure 2 displays the differential time delays of the inner image on the same side of the source, and both images on the opposite side of the source in relation to the primary image (the outermost image on the same side as the source). The differential time delays are always positive. Virbhdara and Ellis \cite{ve00} determined that relativistic images are {\em extremely} demagnified and therefore we did not attempt to study those in this paper. We also studied variation in image positions of all four images as the source position changes. 
\indent
Observation of negative time delay signatures in strong field lensing would imply the existence of SNS. The relationships shown in Figures 2 and 3 and Tables I and II comprise lensing signatures that would be characteristic of SNS lensing for $\nu=0.01$. Finding any lensing signature of an SNS would disprove not only the cosmic censorship hypothesis of Sir Roger Penrose \cite{pen69}, but also a weaker version of cosmic censorship hypothesis by Virbhadra\cite{v09} that essentially states that marginally and strongly naked singularities do not exist in a realistic gravitational collapse.  \\
\indent
It is of great importance to say that in the proximity of a galactic center, there is significant extinction of electromagnetic radiation. Therefore, such observations are not easy. With improved observational techniques in the future, observation of a SNS would be one of the most important successes in astronomy and would also help us study quantum gravity. SNS have the unique ability to be distinguished from black holes observationally, thus making their observation auspicious to disproving CCH. We considered the JNW spacetime that has scalar field. Scalar fields are often treated as dark energy which is believed to make up a majority (around $70\%$) of our universe, therefore effects on lensing due to dark energy are expected\cite{kh13}. Results obtained in this paper might also help in the study of dark energy lensing.

\acknowledgments
We would like to thank our mentor, K. S. Virbhadra, for his unwavering dedication and support. The time he devoted to us is invaluable, and he can not be thanked enough.

\end{document}